\documentclass[runningheads]{llncs}
\usepackage{graphicx}
\usepackage{subfig}
\usepackage[toc,page]{appendix}
\usepackage{amsmath}
\usepackage{mathrsfs} 
\usepackage{bbm}
\usepackage{mathtools} 
\usepackage{amssymb}
\usepackage{float}
\usepackage{todonotes}

\DeclarePairedDelimiter{\norm}{\lVert}{\rVert}

\title{Blind De-Blurring of Microscopy Images for Cornea Cell Counting}
\author{Alon Tchelet \and Stefano Vojinovic \and Leonardo Mussa}
\institute{EPFL\footnote{The project was supervised by Majed El Helou and Professor Sabine Süsstrunk in CS413 at EPFL}, Lausanne, Switzerland}

\begin{document}
\maketitle

\begin{abstract}
Cornea cell count is an important diagnostic tool commonly used by practitioners to assess the health of a patient's cornea. Unfortunately, \textit{clinical specular microscopy} requires the acquisition of a large number of images at different focus depths because the curved shape of the cornea makes it impossible to acquire a single all-in-focus image.
This paper describes two methods and their implementations\footnote{python implementations on: \url{https://github.com/ATchelet/Team4\_Cornea\_Debluring}} to reduce the number of images required to run a cell-counting algorithm, thus shortening the duration of the examination and increasing the patient's comfort. The basic idea is to apply de-blurring techniques on the raw images to reconstruct the out-of-focus areas and expand the sharp regions of the image. Our approach is based on blind-deconvolution reconstruction that performs a depth-from-deblur so to either model Gaussian kernel or to fit kernels from an ad hoc lookup table.

\end{abstract}

% \begin{figure}
%     \centering
%     \includegraphics[width=4cm]{images/EPFLlogo.png}
% \end{figure}

\newpage

\section{Introduction}

\subsubsection{General Problem Description} The cornea is the transparent structure that covers the front portion of the eye and its main function is to focus the incoming light. It is composed by proteins and cells, whose density provides a good indicator of the general health, pharmaceutical exposure, ocular surgical procedures and senescence of the cornea endothelium \cite{Review_cornea}.
For these reasons, the corneal Endothelial Cell Density (ECD) is widely used to diagnose pathological conditions and to asses the overall health of the cornea \cite{cornea_book}.
Currently the most common diagnostic tools used to measure the ECD are confocal microscopy and specular microscopy, the main difference being that confocal microscopes use lasers to illuminate a specific region of the specimen while specular microscopes acquire a \textit{wide-field} image \cite{book-microscopes}.
Confocal microscopy is designed to give a sharp image of the exact plane of focus, as it only captures the light coming from the small area which was lit by the lasers. On the other hand, specular microscopy acquires a whole image of the sample, producing an image which is affected by out-of-focus blur \cite{cornea_book}.

Assessing the ECD of a patient with a \textit{clinical specular microscope} requires the acquisition of multiple images of the patient's cornea at different focal depths, since the curved shape of the cornea and the narrow depth-of-field of the microscope make it impossible to acquire a single all-on-focus image. 
Even if ocular microscopy is a non-invasive procedure \cite{Review_cornea}, a long examination time can cause discomfort to the patient and it is clear that finding a method to reduce the number of required images would benefit both the patient and the practitioner.

\subsubsection{Current Solutions}
ECD is usually performed in an automated or semi-automated way by cell counting software which is often included with the instrument \cite{cornea_book}.
As said before, confocal microscopy is gaining popularity as a diagnostic tool, however it is a relatively slow technique which requires the acquisition of a whole scan of the specimen and  higher levels of light excitation \cite{book_deconvolution-microscopy}.

Deconvolution is a very popular method that seeks to computationally reverse the blurring effects of the microscope. Many different deconvolution algorithms exist, and generally speaking they attempt to reassign the light coming from out-of-focus points to its original location \cite{Quantitative_deconvolution_microscopy}, usually taking as input a 3D stack of images. However, in order to yield good results, most de-blurring algorithm must be tailored to the specific application they were intended for, and require some kind of knowledge about the characteristics of the data they work on \cite{Blind_Image_Deconvolution}.
Other methods leverage chromatic aberration to extend the depth of field of a channel using the information acquired with the other channels \cite{correlation_based_deblurring}. Recently, also in the Deep Learning domain there have some efforts towards the training of neural networks capable of \textit{blind deconvolution} \cite{Ren2019} \cite{Kupyn2018} \cite{Chakrabarti2016}. 
To our knowledge, there have been relatively few attempts to develop an image de-blurring algorithm which can extend the focus of a single 2D image, in the field of corneal microscopy \cite{Depth_Estimation_and_Blur_Removal_from_a_Single_Out-of-focus_Image}. 

\subsubsection{Proposed Solution}
In this paper we present a procedure which allows to extend the focus of a cornea microscopy picture, starting from a single 2D image acquired with a specular microscope.
The goal is to reduce the number of images required during the examination of the cornela endothelium, by enlarging the sharp region of each scan and thus increasing the patient's comfort.  
Our algorithm is based on a space-variant deconvolution, with blind-estimation of Gaussian kernels which are modeled according to the distance from the focal plane. The kernels model is built according to a previous analysis of ground-truth kernels, which is derived from a training set of images. 
The distance from the focal plane is estimated from the magnitude of the defocus blur present in the image, as explained in more detail in Section \ref{sec:impl}.

\section{Theoretical Background}
At its core, the problem we tackle during our research can be reduced to an image restoration problem.
For this reason, it is important to build a good theoretical understanding ans well as a good understanding of the most used algorithms within the field of image restoration. In the following section we provide a brief introduction to the subject, provided with a small review of the standard algorithms and methods used in the field.

\subsection{Image Restoration Overview}
Image restoration is a fundamental problem in image processing, whose applications encompasses a wide range of disciplines. Virtually all disciplines in which images are acquired at difficult environmental conditions, they can benefit from image restoration techniques: medical imagery, astronomy, photography are some examples. \cite{the-Essential-Guide-to-Image-Processing}

Generally speaking, all image restoration algorithms have the goal of improving the quality of an image that was corrupted by noise and whose information or aesthetic content has been degraded. This is usually done by establishing a mathematical model which describes the blurring process. 
A very common degradation model is based on the convolution of the \textit{original image} with the PSF \cite{image_restoration_book}.

\begin{equation}
    b = i \circledast PSF + N
\end{equation}
where, $b$ represents the \textit{blurry image} observable after restitution, $i$ is the \textit{ideal image} which is the scene we desire to capture and $N$ is the noise contribution. Note that for sake of simplicity we will stick to this notation throughout the document and we will denote the Fourier transforms of a variable with an uppercase letter.

The PSF is nothing more than an image describing the projection of a point onto the camera sensor, and its shape and size depend on multiple factors including camera lenses aberrations \cite{axial_chromatic_aberration}, distance of the point from the focus plane and motion of the camera \cite{A_Workingperson_Guide}.  Note that in theory, each point of the original image should be associated to a different PSF, as the projection mechanism depends on the position and on the depth of the original image point. However, for practical purposes neighboring pixels are usually assumed to share the same PSF and it is common practice to partition the image into smaller patches and to assign a PSF to each of them.

This model poses two major challenges to the successful restoration of the original image. Firstly, it is necessary to invert the convolution operation, performing what is known as a \textit{deconvolution}. Secondly, it is necessary to find a way to estimate the PSF of the image to be restored, which has proven to be an arduous task.  
Over the course of the years, different methods have been developed to address this problem and each has its own strength and weaknesses \cite{A_Workingperson_Guide}. A first major distinction can be made between those methods which assume a prior knowledge of the PSF (\textit{non-blind deconvolution}) and those which estimate both the original image and the PSF at the same time (\textit{blind-deconvolution).} 

In the following sections we will describe some of the most common restoration algorithms existing today, starting with a naive example which is rarely used in practice, but nevertheless helps building a good understanding of the principles underpinning deconvolution.

\subsubsection{Inverse Filter}
At a first sight, restoring the original image appears to be a fairly simple task, provided that the PSF is known (we will discuss more in detail methods to estimate the PSF later on).
A naive solution would be to translate the problem into the frequency domain and divide the blurred image transform by the transform of the PSF (hereafter referred to as the OTF, Optical Transfer Function).

\begin{equation}
   \hat{I}(k,l) = \frac{B(k,l)}{ OTF(k,l)}
\end{equation}
where $\hat{I}(k,l)$ denotes the estimate of $I$, which is the Fourier transform of the \textit{original image} $i$.

This technique is known as \textit{inverse-filter} and may appear as an effective and elegant solution, however there is a major problem. The effectiveness of any restoration technique depends on the assumptions that were made creating the model, which is supposed to describe accurately the degradation process. The inverse filter fails to address the problem of noise \cite{image_restoration_book}, which is found in virtually any image and cannot be eliminated. 
The problem arises when the spectral content of the PSF assumes very small values around some frequencies, which is normally the case with real world PSFs. The noise spectrum, on the other hand, is usually flat and dividing it by the OTF amplifies the noise content of the restored image as a result. 

\subsubsection{Wiener Filter}
\label{section_wiener_filter}
The Wiener filter is an improvement of the simple Inverse Filter described above, which takes into account the effects of noise. The underlying idea is simple: the spectra of the blurred image is divided by the OTF and it is weighted by a coefficient which takes into account the PSNR (Peak Signal to Noise Ratio) of the image at that particular frequency. This improves the reconstruction quality, as those parts of the image where noise is preponderant tend to be assigned a low coefficient and contribute less to the reconstruction of the final image. From a  mathematical point of view the Wiener Filter is a \textit{Mean-Square-Error} estimator \cite{image_restoration_book}, i.e. it minimizes the expectation of the square-error:

\begin{align}
  E[|I(k,l)- W(k,l)  B(k,l)|^2] \\
  W(k,l) = \frac{H^*(k,l)}{|H(k,l)^2| \cdot \frac{S_N(k,l)}{S_i(k,l)}}
\end{align}
where, $W$ represents the Wiener coefficient while $S_N$ and $S_i$ represent the Spectral Power Density (the energy at a given frequency) of the ground-truth image f and of the Noise respectively. Note that when the fraction $\frac{S_N(k,l)}{S_i(k,l)}$ assumes a large value (relatively high noise content) the Wiener coefficient assumes a small value, so that the algorithm "ignores" this frequency. The main issue here is that to calculate $S_N$ and $S_i$ knowledge of the \textit{auto-correlation} of the original image and of the noise is required. 
In practice the noise power distribution can be estimated from the characteristics of the camera sensor, while the Spectral Power Density of the image is a bit more challenging. Usually this latter is estimated from a similar image, which yields good results given that most images have a similar power distribution and the Wiener Filter is not too sensitive to the exact power distribution of the image.

\subsubsection{Constrained Iterative Deconvolution}
The problem with the inverse filter approach is that it treats the images as if they were noiseless, forcing the reblurred solution $H \hat{i}$ to be equal to the blurred image {b} \cite{image_restoration_book}. This "hard" constraint introduces a large amount of high frequencies in the reconstructed image $\hat{i}$ deteriorating the result.
A common approach to overcome this problem is to formulate the restoration problem as an optimization problem, i.e. to minimize a cost function through an iterative procedure. 
First an \textit{initial guess} is made, usually applying a simpler algorithm such as inverse filtering. Then the guessed image is reblurred (convolved with the PSF) and a cost function is used to compare the reblurred and acquired images. Lastly, corrections are applied to maximize the match of the two (usually gradient descent \cite{Fourier-Domain-Majed}).
The term "constrained" comes from the fact that usually some constraints are imposed during the restoration process, such as \textit{non-negativity} and low \textit{high-frequency noise} of the restored image.
This is usually done by introducing a so-called \textit{regularization term}, as in the following example:

\begin{align}
   \operatorname*{min}_{\hat{i}}   \{\norm{\textbf{b} - h \circledast \hat{i}} + \alpha \norm{L\hat{i}} \}
\end{align}
where, $\alpha \norm{L\hat{i}}$ represents the regularization term: in this example $\alpha$ is a constant and $L$ is a \textit{high-pass filter} written in matrix form, note that the regularization term could be substituted by any other suitable function which incorporates an \textit{a-priori} knowledge about the original image.

\subsubsection{Blind Deconvolution}
In those cases when the microscope PSF is difficult to obtain or too imprecise to yield good results, it is possible to estimate the ground-truth image and the PSF at the same time, performing what is known as a \textit{blind-deconvolution}.
This is inherently harder than simple deconvolution because there exists an \textit{infinite} number of pairs (estimated image and blur kernel) which solve the problem, i.e. it is an \textit{ill-posed} problem \cite{Understanding-Blind-Deconvolution}. So far, the prominent strategy is to include into the algorithm some a-priori knowledge about the image, in order to prune nonsensical solutions from the solution space \cite{paper-blind}. 

Usually, the \textit{a-priori} knowledge about the data is incorporated into a cost function, which is minimized to obtain estimates of the original signals \cite{Blind_Image_Deconvolution}.
Some of the most commonly used constraints for the blur kernel include \textit{smoothness}, \textit{non-negativity} and \textit{symmetry} to cite a few examples \cite{the-Essential-Guide-to-Image-Processing}, and if well chosen they can transform the problem into a well-posed one. Another successful strategy which was applied in recent years is to constrain the estimated image taking into account natural image statistics, which are computed from large data sets \cite{Understanding-Blind-Deconvolution}.
This type of statistical \textit{priors} have to be handcrafted and target a specific category of images, however recently some efforts have been made to construct \textit{priors} with deep learning methods such as Convolutional Neuronal Networks (CNN) \cite{CNN-prior}.

Blind deconvolution can be further classified into two categories, based on the manner in which the original image is estimated \textit{A priori}. Blind deconvolution estimates the blur kernel separately and then applies a deconvolution algorithm to produce the image estimate, in a two steps process. On the contrary,\textit{ joint blind deconvolution} performs the two tasks in one single step. \cite{Blind_Image_Deconvolution}
Blind deconvolution problems, can be solved with any of the well-established algorithms used to solve optimization problems, such as \textit{successive approximations} to name an example. The challenge is to pick the exact algorithm, constraints and parameters which best suit the set of images to be deblurred, a procedure which often requires a good deal of experience and a trial and error approach.

\section{Implementation}
\label{sec:impl}
As discussed in the introduction, the purpose of our work is to extend the sharp area of a clinical image of the cornea, so that less images are required to run a cell-counting algorithm.
The cornea is a curved structure and therefore the images are largely out-of-focus, with a fast transition between the sharp regions and the blurred ones. Given the narrow depth-of-field of clinical microscopes, some areas of the images are so blurred that it appears unreasonable to attempt a restoration of these regions. We chose to focus our efforts on the transition area between the sharp and the blurry region, where the image is too degraded for the cell-counting algorithm to work but the information content can still be recovered.

\subsection{General Description of the Algorithm}
The method that we propose is based on \textit{blind-deconvolution} with an \textit{a-priori} estimation of the kernels. The kernels estimation was performed according to two different methods so to assess which one leads to the best restoration. The first method models Gaussian distributed kernels while the second relies on a look up table, both approaches fit the kernels depending on depth blur.  The information necessary to build the models was obtained by analysis of the ground-truth kernels, which were obtained by comparison of a sharp and blurry image of the same patch, as described in more detail later on. 

Deconvolution assumes the PSF to be space-invariant across the whole image, however this was obviously not the case since there is a wild variation in depth across a single image.
For this reason, we partitioned the image into a grid of small overlapping patches, which were assumed to have similar depth, and we estimated the depth of each patch by measuring the amount of defocus-blur. The patches were then summed and averaged to have a full map of the estimated depth-blur in that area.

Finally, we deblurred each patch with a Wiener restoration algorithm and stitched the results together to reconstruct the original image. 
It is important to note that the choice of the patch size has an important impact on the quality of the results, as very large patches may break the assumption of relatively constant depth and conversely, too small ones may suffer from increased boundary effects or translate badly into frequency space due to their little data content.
Over the course of our research we found a patch size of 64x64 to be a good compromise between resolution and information content of the patches.

In the following section we describe more extensively the fundamental steps of the algorithm.

\subsection{Pre-processing}
The ground-truth kernel estimation step requires the comparison of a blurry patch with its sharp corresponding image. In order to perform this comparison with ease, we generated a single \textit{all-sharp} image, composed by the union of the sharp regions of multiple images (stack fusion).
The pre-processing is divided in two steps namely the image registration and the stack fusion.
The first step aligns all the images representing the same subject at different depth-focus, while the second step can be described as a procedure which selects the sharp areas of each image and stitch them together to generate an all-focused image. 
\begin{figure}%
    \centering
    \subfloat[focused]{{\includegraphics[width=4cm]{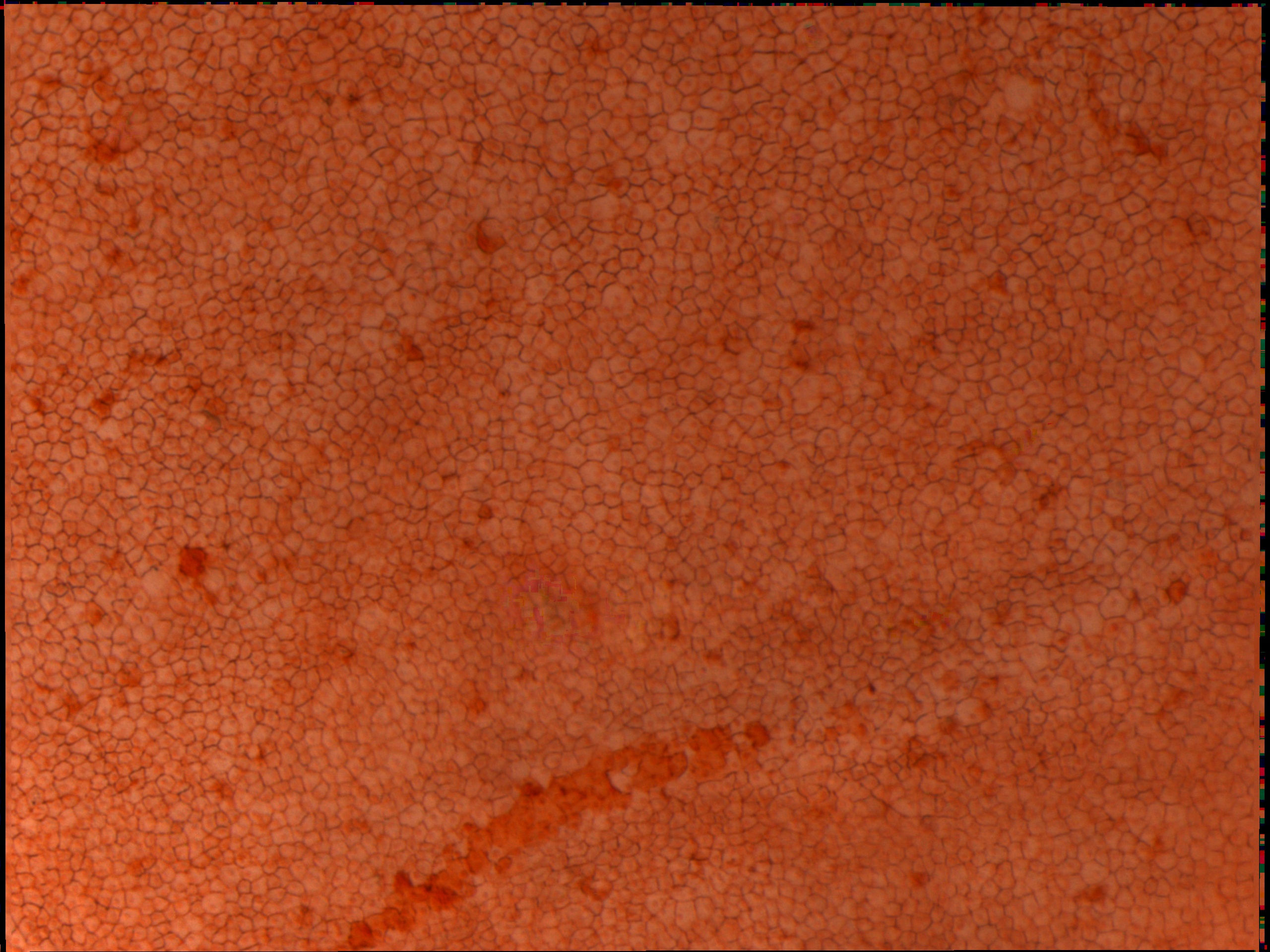} }}%
    \qquad
    \subfloat[blurry]{{\includegraphics[width=4cm]{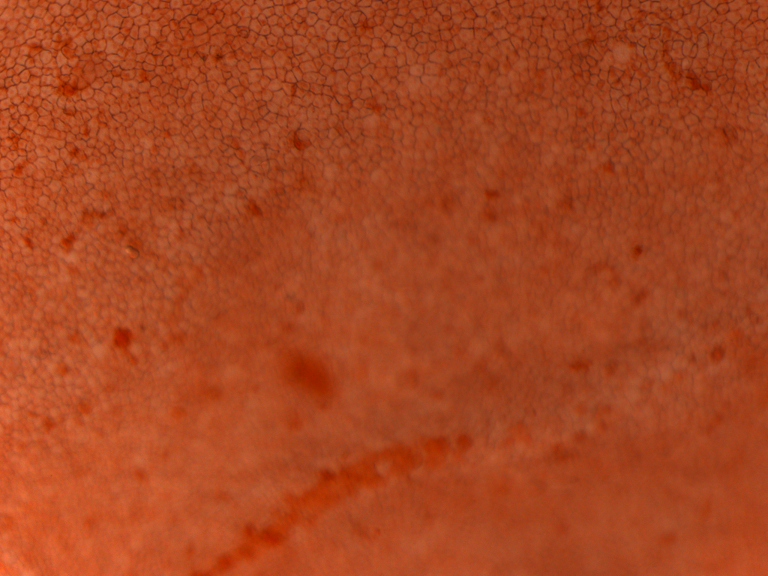} }}%
    \caption{A focused image obtained by stack-fusion (a) compared to a single blurry image of the stack (b).}%
    \label{fig:example}%
\end{figure}
By performing the aforementioned steps it is possible to reconstruct the ground-truth image or \textit{ideal image}. The reconstruction of the \textit{ideal image} is a crucial steps since it allows to perform comparisons between the \textit{blurry images} received from the microscope an the \textit{ideal image} which is the ultimate target of the \textit{blind-deconvolution} process.

\subsection{Ground-truth Kernels Estimation}
The kernels estimation is a fundamental step since it allows to understand better the properties of the blur affecting the images acquired by the microscope. At first sight it can be observed that the \textit{defocus-blur} is not constant throughout the entire image but it rather increases in magnitude as we move further from the in focus region, as it can be seen in \ref{fig:example}b. This intuition is the starting point for the  blur estimation with a mathematical models. Such model relies on two inputs, the \textit{blurry image} and the \textit{ideal image} which are both portioned partitioned in patches of 64x64 pixels so to have local estimations of the blur. The size of 64x64 has been empirically found as a good estimation constant local blur.

When performing the kernel estimation the cost function to be minimized presents an L2 regularizer and it can be formulated as follows:
\begin{equation}
   \operatorname*{minimize}_{h_{x,y}}   \{ \norm{b_{x,y} - h_{x,y} \circledast i_{x,y}}^2 +  \norm{\lambda h_{x,y}}^2 \}
\end{equation}
where, $b_{x,y}$ and $i_{x,y}$ refer to the blurry and sharp patches of coordinates (x,y) and $h_{x,y}$ is the blur kernel associated to them.
Note that this objective function can be rewritten in a simpler form replacing the convolution operation with a matrix multiplication. This is possible rearranging the pixels of the image $i_{x,y}$ into a column vector and those of the kernel $h_{x,y}$ into a suitable circulant matrix $H$ as described in \cite{Fourier-Domain-Majed}.
\begin{equation}
   \operatorname*{minimize}_{h_{x,y}}   \{ \norm{b_{x,y} - H_{x,y} i_{x,y}}^2 + \lambda \norm{ H_{x,y}}^2 \}
\end{equation}

This is a convex problem (Frobenius norm) which admits a unique solution and which can be solved in closed form by setting the derivative with respect to the image $i_{x,y}$ to zero  \cite{Fourier-Domain-Majed}. In our case the optimization variable is the kernel $H_{x,y}$ and not the deblurred image $i_{x,y}$, which is known.
\begin{equation}
    \frac{\partial[ \norm{b - H i}^2 + \lambda \norm{ H}^2]}{\partial i}  = 2\lambda H^T (b - H i)   = 0
    \label{eq:majed}
\end{equation}
\textbf{Remark:} for convenience we will drop the pedix $_{x,y}$. For now on, variables will be referred to a generic patch of coordinates (x,y).\\

Starting from equation (\ref{eq:majed}), an efficient closed form solution leveraging some useful properties of circular matrices can be performed so to compute the ground-truth kernels \cite{Fourier-Domain-Majed} . 

The kernel estimation is performed on the green channel of the images, since it has been assessed that such channel is the one conveying the most relevant information. The following analysis is described in a qualitative way since its purpose is to illustrate at high level how the blur affects the microscope images.
\begin{figure}[h]
    \centering
    \subfloat[blurry image]{{\includegraphics[width=4cm]{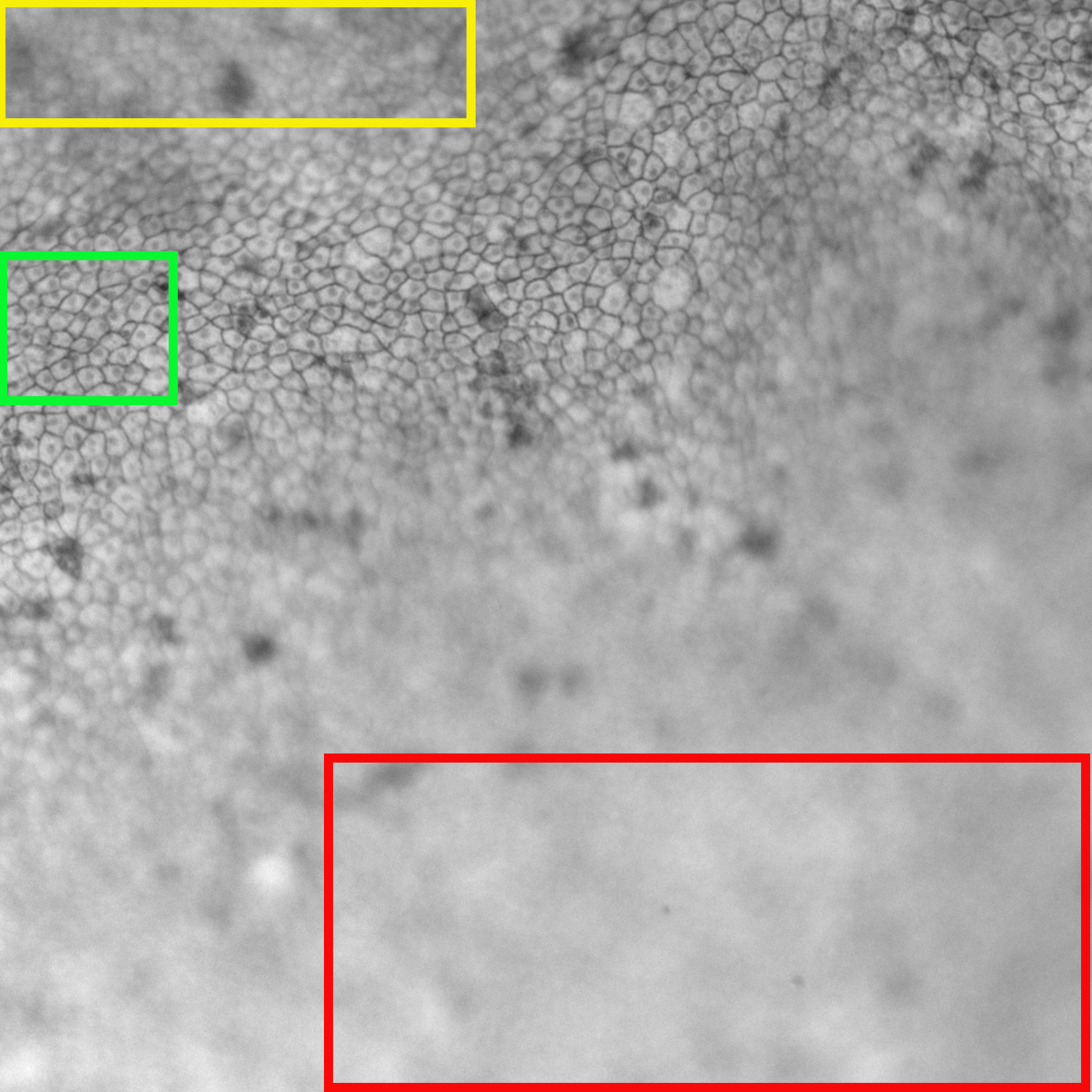} }}%
    \qquad
    \subfloat[kernel map/blur map]{{\includegraphics[width=4cm]{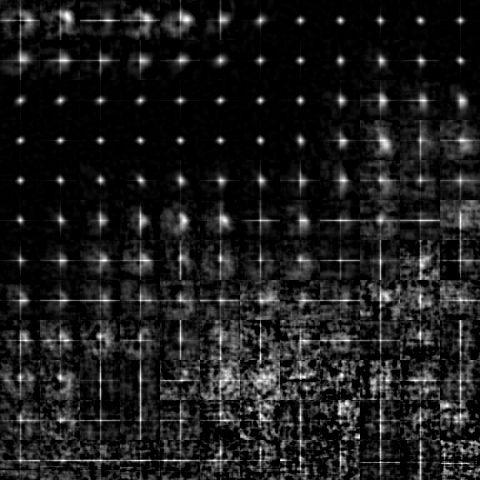} }}%
    \caption{The green channel of the image obtained from the microscope(left) and the ground-truth kernels corresponding to each 64x64 patch (right). The colored rectangles highlight samples of areas with different amount of blur. The green box samples a part of a sharp area with negligible amount of blur, the yellow box samples a part of slightly blurry area and the red box samples a part of highly blurry area.}%
    \label{fig:bl_km}%
\end{figure}

The kernels quality across the entire range is strongly dependent on the blur level on which the estimation is performed. As it can be observed in Fig. \ref{fig:bl_km} the areas that are in the sharp region (green box) do present a PSF that tends to resemble a Gaussian kernel. However, as soon as we move towards areas that are on the boundary of the sharp region (yellow box), the kernel estimates do present a form of structure but without a recurrent shape. Finally, in the most blurry areas (red box) the estimation is so poor that the kernel do not present a structure and they just have a noisy form. Therefore, this analysis suggests that only the slightly blurry areas (yellow box) can be recovered while areas strongly affected by blur (red box) cannot be recovered with this kind of estimation. 

\begin{figure}[H]
    \centering
    \includegraphics[width=12cm]{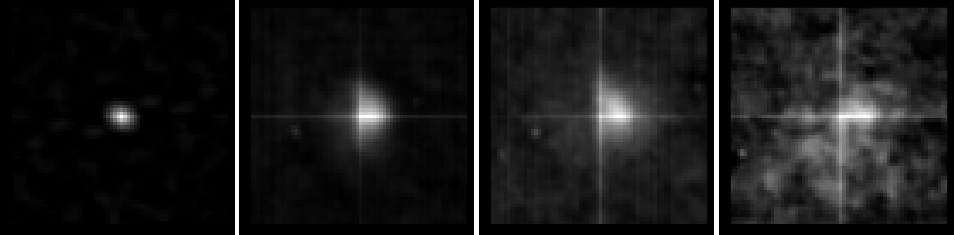}
    \caption{Estimated Ground-Truth Kernels. They are in ascending order of blur from left to right.}
    \label{fig:blur_kernels}
\end{figure}

\begin{figure} [H]
    \centering
    \subfloat[blurry image]{{\includegraphics[width=3.5cm]{images/Ori_C_1Rec.jpg} }}%
    \qquad
    \centering
    \subfloat[ideal image]{{\includegraphics[width=3.5cm]{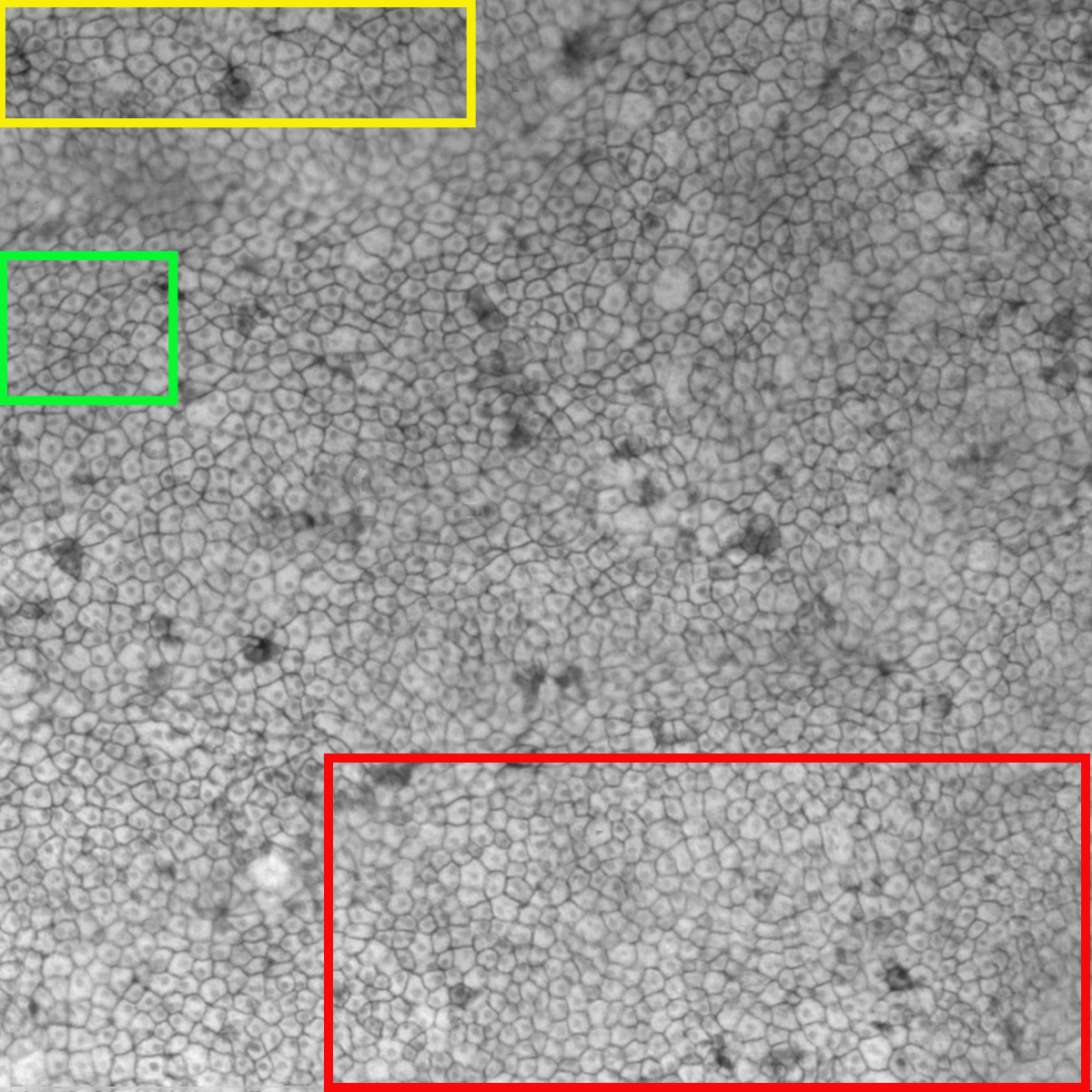} }}%
    \qquad
    \subfloat[restored image]{{\includegraphics[width=3.5cm]{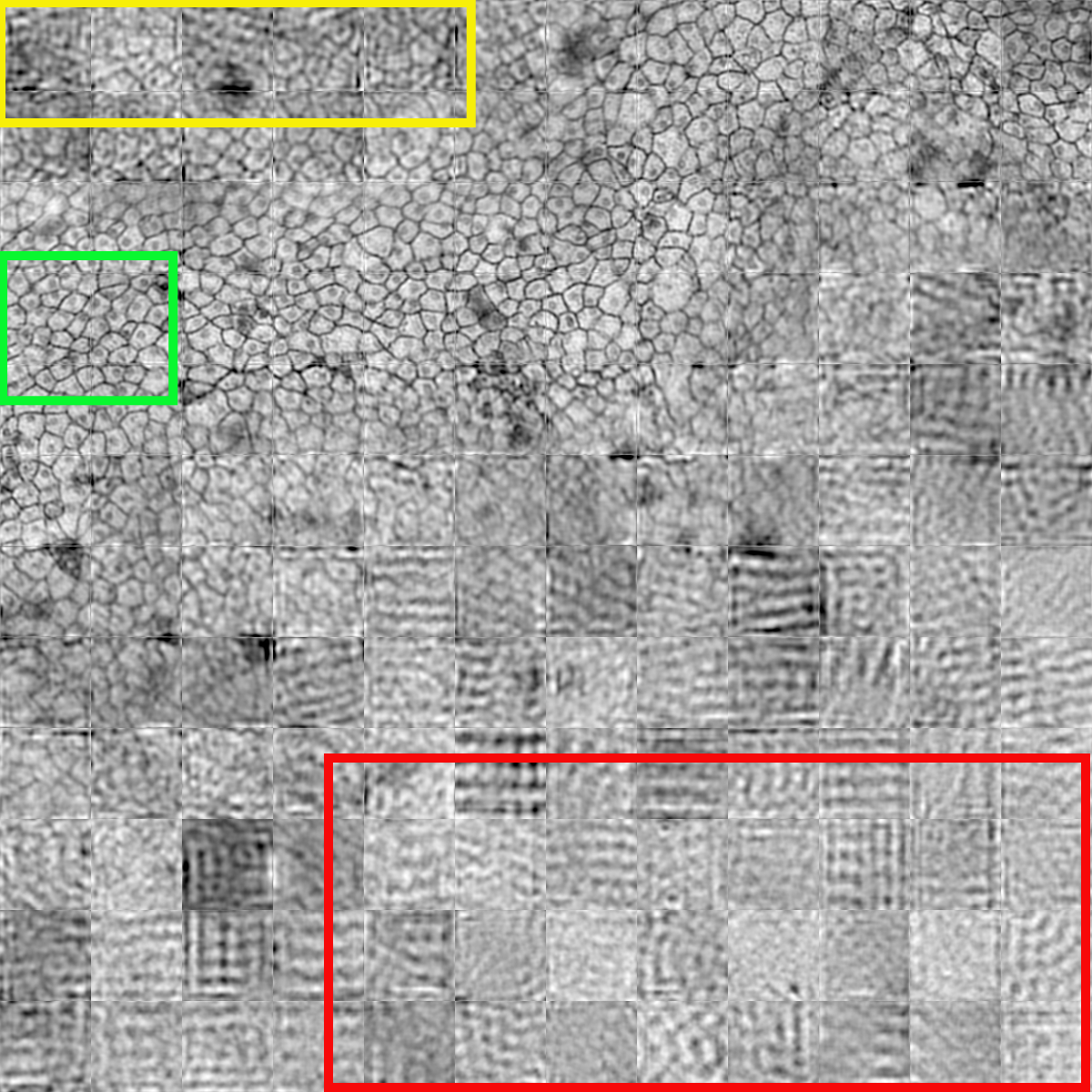} }}%
    \caption{The input image from microscope (a) The ideal image all in focus (b) and the restored image(c).}%
    \label{fig:ii_rI}%
\end{figure}

In order to validate the hypothesis formulated after observing Fig. \ref{fig:bl_km}, the \textit{blurry image} is deconvolved with the \textit{kernel map} and then compared with the \textit{ideal image}, as shown in Fig. \ref{fig:ii_rI}. At first glance it can be observed that the sharp areas (green box) are maintained intact while the strongly blurry (red box) are just dominated by artifacts. However, the areas that were slightly affected by blur (yellow box) do convey more information in the \textit{restored image} than in the 
\textit{original image}.
It can be concluded that the algorithm for estimating kernels operates in a satisfactory way on areas undergoing a slight amount of blur while it does not estimate correctly heavily degraded areas. By observing the distribution of the kernels in the recovered regions (yellow and green box), we propose two approaches. The first one, is about approximating the observed kernels with  Gaussian distributed kernels. The second one, is about creating a look up table made of the observed kernels.  The next section provides an extensive description on how such kernels are used so to build a blind deconvolution algorithm.
 
\subsection{Depth Map Estimation}
The literature offers many different methods to estimate depth from defocus-blur, for example using the gradient across edges \cite{Defocus-from-edge}, reblurring an edge with a Gaussian kernel \cite{Low-cost-blur-estimator} or a low-pass filter \cite{blur-level} and analyzing the difference, or by comparing  local contrast with gradients magnitude (local contrast prior) \cite{depth-from-local-contrast}. As a general rule, the good success of any of these methods depends on the amount of textures and visual features present in the image, which should be as rich of details as possible \cite{depth-from-local-frequency}.

To estimate the amount of blur the Crété method was used. The method uses the fact that a blurry area has a far smaller variation when being re-blurred than a sharp image being blurred. In the method, the intensity difference between neighboring pixels is calculated for both the original image and a blurred version on the horizontal and vertical axes. Then, the absolute differences of each focused and blurred pixel neighborhoods is calculated. With those differences matrices the variation between the focused and blurred on each axis are calculated. Sums of the coefficients in each focused and the variation matrices of both axes are taken and the normalized difference between them on each axis is the blur level of the axis. The maximal value between the two represents the blur level of a segment \cite{blur-level}.

\begin{figure}[H]
    \centering
    \includegraphics[width=12cm]{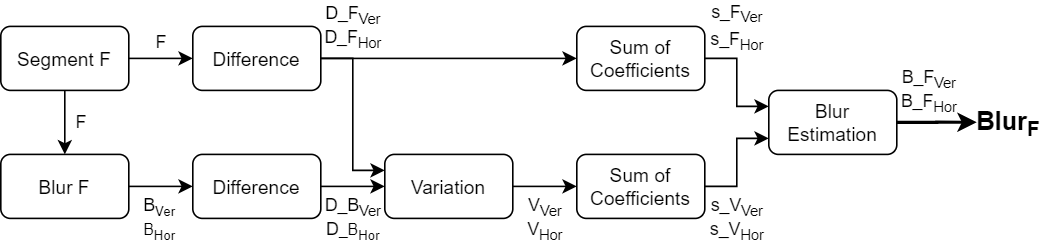}
    \caption{Diagram of the blur level estimation algorithm.}
    \label{fig:blur_alg}
\end{figure}

In our implementation the blur is calculated on square segments and is overlapped in order to get a continuous picture of the blur level across the image. The larger the segment width is the more smooth the resulting map. And the larger the overlap between the segments the more granulated the map is.

\subsection{Blind Gaussian Kernel Estimation}
As discussed before, \textit{blind-deconvolution} is a strongly ill-posed problem as it is necessary to estimate the blurring kernel and the original image at the same time. Fortunately, in the case of motion-blur or out-of-focus blur it is possible to employ parametric kernels which are derived from theoretical considerations on the blur mechanism. In the specific case of defocus-blur it is common to describe the PSF with a \textit{gaussian-model} or with an uniform disk function. In mathematical symbols the gaussian PSFs can be described by:

\begin{equation}
    \label{equation_PSF_gaussian}
    PSF_{gaussian}(d) = \frac{1}{2\pi\sigma} exp(-\frac{d^2}{2\sigma^2})
\end{equation}
where, $d$ stands for the distance from the center of the PSF and $\sigma$ denotes the standard deviation of the gaussian function. Note that the parameter $\sigma$ depends on the depth of the patch, which can in turn be estimated from the amount of blur of the region under consideration.

Once the depth map is created it is possible to use  equation (\ref{equation_PSF_gaussian}) to associate a Gaussian kernel to each patch, since the parameter $\sigma$ can be estimated from the depth of the region under consideration with the following expression:

\begin{equation}
    \label{equation_sigma}
   \sigma = 50 \cdot Blur
\end{equation}
where, $Blur$ is the percentage value estimated with \cite{blur-level} and $50$ is an empirical value that was found to yield the best results.

\subsection{Kernels Lookup Table}
A second approach implemented was to create a lookup table according to the kernels estimated from the ground-truth image. The lookup table is constructed by estimating the kernels of the blurred images that generated the ground-truth image and relate them to specific blur levels. All kernels related to a specific blur value are then averaged and missing blur values could be interpolated from the neighboring values.

\subsection{Image Restoration with Deconvolution}
The individual patches were deblurred with a Wiener restoration, which was introduced in section \ref{section_wiener_filter}, and stitched toghether to form a single deblurred image. The specific implementation that was used is the function \textit{skimage.restoration.wiener} from the \textit{skimage} Python package which is based on \cite{paper-wiener-python}.
This function assumes the following (standard) model for the data degradation mechanism:

\begin{equation}
    b = i \circledast PSF + N
\end{equation}
where, N is the noise, $i$ is the unknown original image and $b$ is the blurred image.

the algorithm then performs the following operation to deblur the image
\begin{equation}
	\hat{i} = \mathscr{F}^{-1} ({|OTF|}^2 + \lambda {|\Lambda_D|}^2) \cdot OTF^{-1} \cdot B
\end{equation}
where, $\hat{i}$ is the estimated original image, $OTF$ is the Fourier transform of the PSF (Optical Transfer Function), $B$ is the Fourier transform of the blurry image $b$ and $\Lambda_D$ is a filter to penalize the chosen restored image frequencies. Note that by default a Laplacian filter is used which penalizes high frequencies. The parameter $\lambda$ is used to tune the balance between the regularization factor and the data, which tends to amplify the noise and consequently the higher frequencies of the spectrum. 
Note that to obtain the best results it is necessary to customize the parameter $\lambda$ depending on the amount of noise present in the image, after empirical evaluation we adopted a value of 0.1.

\section{Results}

\subsection{Comparison with Other Methods}
\begin{figure}[H]
    \centering
    \subfloat[Original]{{\includegraphics[width=3.5cm]{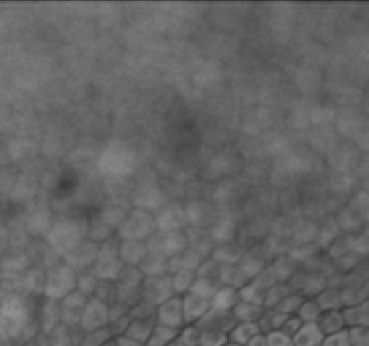} }}%
    \subfloat[Deblurred]{{\includegraphics[width=3.5cm]{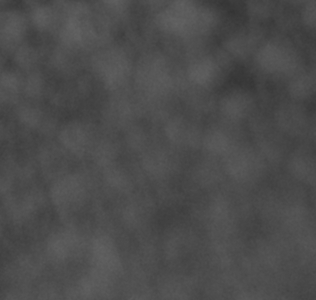} }}%
    \caption{The results of a \textit{regularized inverse filter} deconvolution which uses a theoretical PSF generated according to an optical model of a microscope.}
    \label{fig:ij}%
\end{figure}

 Blind deconvolution is an intrinsically hard problem, which requires a good deal of tuning and usually targets a specific type on data on which the priors are tuned. For these reasons it was not an easy task to find software that could perform a true \textit{blind-deconvolution}, as most open-source material was either of very poor quality, or targeting different problems such as 3D deconvolution or aesthetic image enhancement.
 Nevertheless, during the course of our research, we applied some de-blurring algorithms (regularized inverse filter \cite{deconvolutionlab2} and others) on our data to assess the quality of our result and set a reference for our expectations.
The best resource we could find is the \textit{deconvolutionLab2} tool developed by the \textit{Biomedical Image Group} at EPFL \cite{deconvolutionlab2}.
This software does not perform a truly blind-deconvolution, as it requires a PSF to work, but it includes a tool which allows to generate theoretical PSFs for different types of microscopes. 
Following in Fig.\ref{fig:ij} an example of a deconvolution performed with Imagej is shown.

\subsection{Results of Our Method}
To compare the results of the two de-blurring methods implemented in the project we used three different ways: (1) Visual comparison, (2) Blur map comparison, and (3) Metrics comparison. As blur it a tricky task to compare images based on their blur, the visual examination could be useful sometimes. Comparing the blur maps can reveal the change in the blur level on the image before and after the deconvolutions. And finally image metrics can provide some numerical evaluation on the results. A cell count comparison between \textit{blurry image} and \textit{restored image} would lead to an optimal metric that would quantify the benefits brought by the blid-deconvolution algorithm proposed

\subsubsection{Visual Comparison:}
\begin{figure}[H]
    \centering
    \subfloat[Original]{{\includegraphics[width=3.5cm]{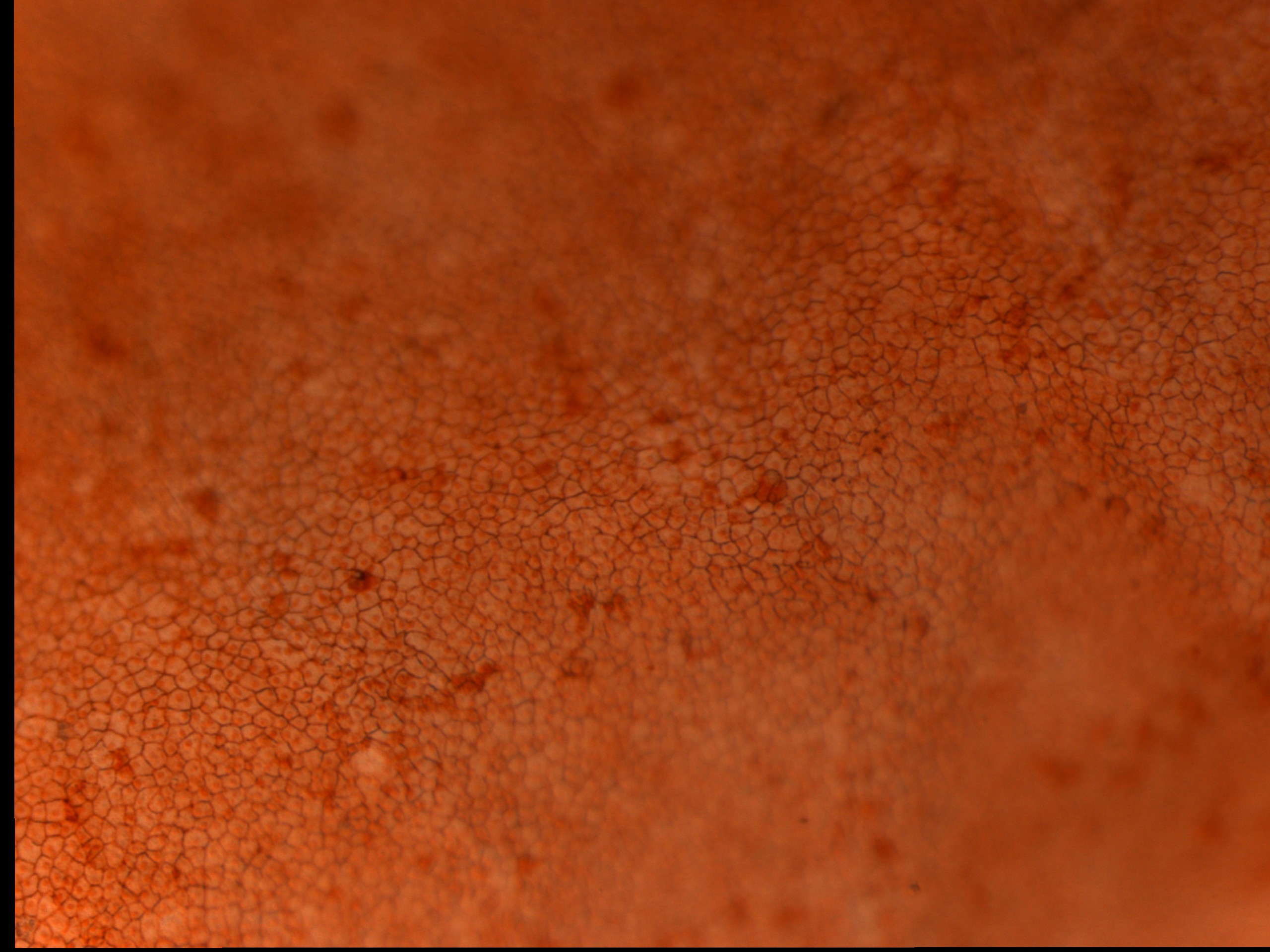} }}%
    \subfloat[Gaussian]{{\includegraphics[width=3.5cm]{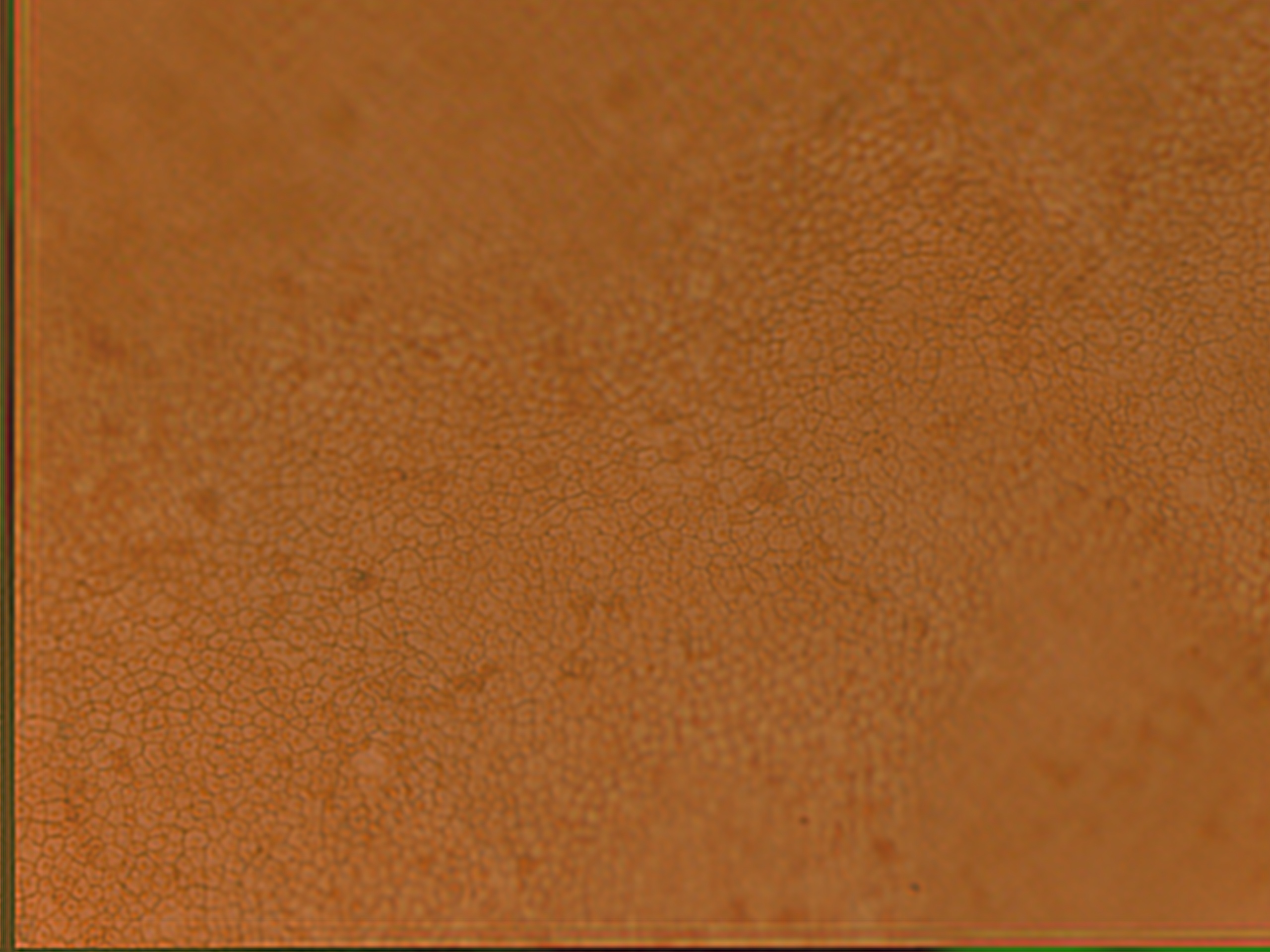} }}%
    \subfloat[Lookup Table]{{\includegraphics[width=3.5cm]{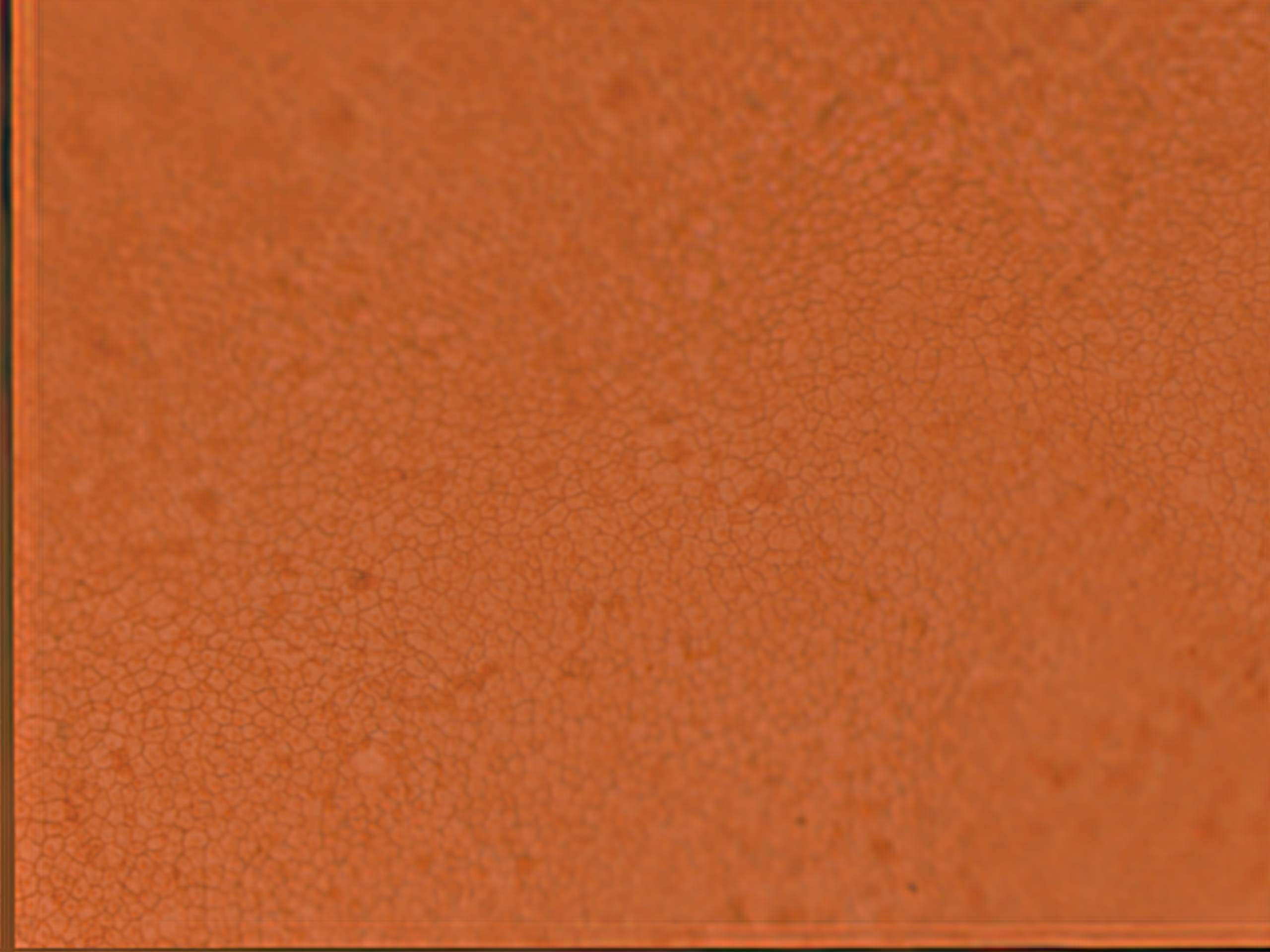}}}
    \caption{Visual comparison of the results using both methods compared to the original image.}
    \label{fig:gausVdeconv}
\end{figure}

As can be seen in Fig. \ref{fig:gausVdeconv}, both methods seem to have slightly expanded the focused area a bit. But it is hard to see to what extent. Despite both methods, and the Gaussian kernels method in particular, affected the color of the resulting image, we can still say that the Gaussian Kernels method have performed slightly better.

\subsubsection{Blur Map Comparison:}
\begin{figure}[H]
    \centering
    \includegraphics[width=10cm]{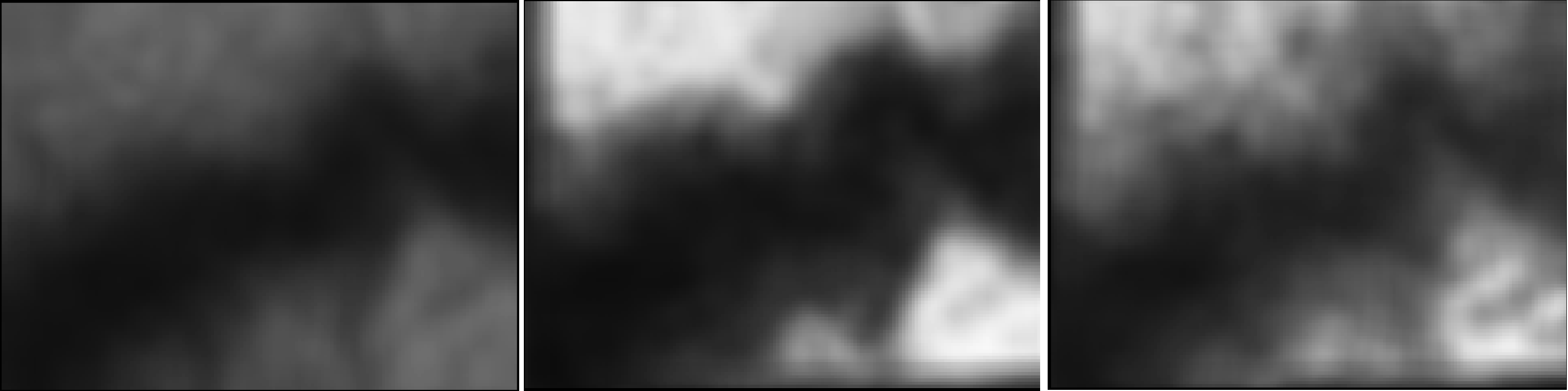}
    \caption{The blur level of the original image (left), the deblurred image using the Gaussian kernels (center), and the deblurred image from the kernels lookup.}
    \label{fig:res_BlurMapImage}
\end{figure}

From Fig. \ref{fig:res_BlurMapImage} we can see that the Estimated Gaussian kernels have reduced the blur to a further point and by a larger factor. The Kernels \textit{lookup table} method has also succeeded in extending the clear area and seems like it extended it to roughly the same distance the Gaussian Kernels method did. Although the blur level in the edge is higher than the one of the other method. Interesting to see is that both methods have also increased the amount of blur in the deep blur areas of the image. 

\subsubsection{Metrics Comparison:}
Both Mean Square Error (MSE) and Peak Signal-to-Noise Ratio (PSNR) metrics were tested, but unfortunately they are both incapable of giving a valuable result. Neither of the methods have any representation of how well cell edges are reconstructed because sharpness is net evaluated. Moreover, measurements of the metrics do not generate useful comparison. The MSE metric is based on the difference between each pixel couple in the sharp and reconstructed images. Therefore noise, blur and shift of values all add to the error although the result might still better. Hence it shall not be used. The PSNR metric is based on the ratio between the square of the maximal intensity and the MSE. Thus it is not only affected by the issues of MSE but also the inner cell values which are not useful as an indicator in our case. Therefore, results of those metrics were exempt from the paper.

\section{Conclusions}
During the project we examined the topic of image restoration and developed an algorithm capable of extending the focused area of \textit{blurry images} for medical use.
To do so, different methods of depth map estimations were examined and eventually the Crété Method was chosen.
Then for the de-blurring process two methods were experimented with: (1) Estimating Gaussian Kernels according to the depth, (2) creating a lookup table of kernels from a focused image relating kernels to blur levels. The first method could be done from a single image after estimating the ratio between the Gaussian's standard variation and the blur level. While the second method needs a pre-existing image set to create a reconstructed ground-truth image to create the lookup table.
Both of the methods did not manage to solve the problem completely, but only improve the margins on the lower blur area. Overall it seems that the estimation of Gaussian kernels dependent on the blur level is the favorable method as it does not depend on a ground-truth image and its result were slightly more promising.

On a final note, the project's work shows that both kernel estimation methods have potential and should be further researched. The implementations are accessible to the public on GitHub and we invite readers to test them and improve upon them.

\subsubsection{Future Work:}
In a continuation to the previous paragraph, it seems that the centers of the PSF kernels are shifted according to their direction and position compared to the field of focus. Hence, the distance and direction of a pixel from the sharp region could be possibly used to further improve the final results. Additionally, we would recommend re-exploring the depth map to evaluate what side of the sharp area is the one closer and which is the one farther from the camera and try to adapt the kernels accordingly. \cite{depth_ambiguity} proposed a  method to solve this depth ambiguity, by exploiting the fact that light of different wavelengths is focused on slightly different focal planes. A multi-channel analysis of the data would then provide enough information to discriminate between points lying on either sides of the focal plane.

\end{document}